\documentclass[11pt,twoside,twocolumn]{article} % **************** setup for double line spacing review
\usepackage[super,sort&compress,comma]{natbib}
\usepackage[version=3]{mhchem}
\usepackage{ctable}
\usepackage{times,mathptm}
\usepackage{sectsty}
\usepackage{balance}
\usepackage{hyperref}
\usepackage{amsmath}
\usepackage{amsfonts}
\usepackage{amssymb}
\hypersetup{colorlinks=true}
\usepackage{svn-multi}
\usepackage{datetime}
\usepackage{graphicx} %eps figures can be used instead
\usepackage{lastpage}
\usepackage[format=plain,justification=raggedright,singlelinecheck=false,font=small,labelfont=bf,labelsep=space]{caption}
\usepackage{fancyhdr}
\usepackage{nomencl} % turn on nomenclasture list
\usepackage{enumerate}
\usepackage{threeparttable}
\usepackage{booktabs}
\usepackage[nomarkers,nolists]{endfloat}

\newcounter{todocounter}

\newcommand{\degree}{\ensuremath{^\circ}} %type degree for degree symbol

\def\mys#1{{\mbox{\tiny{#1}}}}    % font for capital subscripts

\def\dls{\collrad_{\mys{h}}}     % Hydrodynamic radius of particle
\def\collrad{R}                  % radius of colloidal particle
\def\etac{\eta_{\mys{S}}}        % colloid volume fraction
\def\poly{C_{\mys{v}}}           % Coefficient of variation (Polydispersity)
\def\W{W_{\mys{IM}}}            % wt% of ILM

\def\kB{k_{\mys{B}}}            % Boltzmann constant
\def\kBT{\kB T}                 % Thermal energy
\def\Na{N_{\mys{A}}}            % Avogadro's constant
\def\lB{\ell_\mys{B}}    		% Bjerrum length
\def\b{b}    					% Gouy-Chapman length
\def\Mann{\xi}					% Manning parameter 
\def\ms{\lambda}		% Manning parameter for spheres
\def\mc{\Mann_{\mys{c}}}		% Manning parameter for cylinders
\def\m0{\mu_{0}}				% Electrophoretic mobility of sphere of radius lb and elementary charge
\def\Zstr{Z}					% Structural charge on sphere
\def\Zeff{Z_{\mys{eff}}}		% Effective charge on sphere
\def\Zacc{Z_{\mys{acc}}}		% Accumulated charge
\def\nsalt{n_\mys{salt}}			% salt concentraion
\def\debye{\kappa^{-1}}			% Debye length
\def\l{\ell}					% linear backbone charge density
\def\RWS{\collrad_{\mys{WS}}}				% Radius of cylindrical Wigner-Seitz cell
\def\WS{\collrad_{\mys{WS}}}				% Radius of spherical Wigner-Seitz cell
\def\epr{\epsilon_{\mys{r}}}	% dielectric constant
\def\RF{\collrad_{\mys{F}}}		% Imai and Oosawa defined radius at which field exceeds critical value
\def\RI{\collrad_{\mys{I}}}		% Belloni radius at which charge distribution has inflection point

\nomenclature[W]{$\W$}{wt\% of ILM in particle}
\nomenclature[dls]{$\dls$}{Hydrodynamic radius of particle}
\nomenclature[a]{$\collrad$}{Radius of colloidal particle}
\nomenclature[Cv]{$\poly$}{Coefficient of variation (Polydispersity)}
\nomenclature[mu]{$\mu$}{Electrophoretic mobility}
\nomenclature[m0]{$\m0$}{Electrophoretic mobility of sphere of radius $\lB$ and elementary charge $e$}
\nomenclature[etac]{$\etac$}{Colloid volume fraction}

\nomenclature[Na]{$\Na$}{Avogadro's constant}
\nomenclature[kB]{$\kB$}{Boltzmann's constant}
\nomenclature[kBT]{$\kBT$}{Thermal energy}
\nomenclature[beta]{$\beta$}{ $1 / \kBT $}

\nomenclature[alpha]{$\alpha$}{Fraction of counterions condensed}
\nomenclature[lB]{$\lB$}{Bjerrum length}
\nomenclature[l]{$\l$}{Backbone charge density}
\nomenclature[b]{$\b$}{Gouy-Chapman length}
\nomenclature[Rtilde]{$\ms$}{Manning parameter for charged spheres = $\collrad / \b$}
\nomenclature[xip]{$\Mann$}{Manning parameter for charged cylinders = $\lB/ \l$}
\nomenclature[xipstar]{$\mc$}{Critical value of Manning parameter for condensation}
\nomenclature[sigma]{$\sigma$}{Charge density of surface is $e \sigma$}
\nomenclature[epsilon]{$\epsilon_{0}$}{Permittivity of free space}
\nomenclature[epr]{$\epr$}{Dielectric constant}
\nomenclature[y]{$y$}{Reduced electrostatic potential}
\nomenclature[psi]{$\psi$}{Electrostatic potential}
\nomenclature[rho0]{$\rho_{0}$}{Reference number density}
\nomenclature[rho]{$\rho$}{Counter-ion density}
\nomenclature[Z]{$\Zstr$}{Structural charge on sphere}
\nomenclature[Zeff]{$\Zeff$}{Effective charge on sphere}
\nomenclature[Zstar]{$Z^{*}$}{Critical charge on sphere for condensation}
\nomenclature[Zacc]{$\Zacc$}{Accumulated charge inside cell}
\nomenclature[nsalt]{$\nsalt$}{Salt concentration}
\nomenclature[kappa]{$\debye$}{Debye length}
\nomenclature[RWS]{$\RWS$}{Radius of cylindrical Wigner-Seitz cell}
\nomenclature[Rc]{$\WS$}{Radius of spherical Wigner-Seitz cell}
\nomenclature[RI]{$\RI$}{Belloni radius at which charge distribution has inflection point}
\nomenclature[RF]{$\RF$}{ Imai and Oosawa defined radius at which field exceeds critical value}
\nomenclature[Rm]{$r_{\mys{M}}$}{Cross-over radius}
\nomenclature[Xi]{$\Xi$}{Coulomb coupling constant}
\nomenclature[P]{$P(r)$}{Fraction of ions within sphere of radius $r$}
\nomenclature[Sigma]{$\Sigma$}{Conductivity}
\nomenclature[v]{$v$}{Colloidal velocity}
\nomenclature[E]{$E$}{Electric field}
\nomenclature[mu]{$\mu$}{Electrophoretic mobility}
\nomenclature[eta]{$\eta$}{Viscosity}

\begin{document}

\graphicspath{{../figures/}}

\thispagestyle{plain}
\fancypagestyle{plain}{
%\fancyhead[L]{\includegraphics[height=8pt]{headers/LH.pdf}}
%\fancyhead[C]{\hspace{-1cm}\includegraphics[height=20pt]{headers/CH.pdf}}
%\fancyhead[R]{\includegraphics[height=10pt]{headers/RH.pdf}\vspace{-0.2cm}}
\renewcommand{\headrulewidth}{1pt}}
\renewcommand{\thefootnote}{\fnsymbol{footnote}}
\renewcommand\footnoterule{\vspace*{1pt}%
\hrule width 3.4in height 0.4pt \vspace*{5pt}}
\setcounter{secnumdepth}{5}

\svnidlong
{$HeadURL: file:///home/cppb/Dropbox/Electrophoresis-paper/Paper/repo/trunk/CCT.tex $}
{$LastChangedDate: 2013-10-02 16:24:08 +0100 (Wed, 02 Oct 2013) $}
{$LastChangedRevision: 119 $}
{$LastChangedBy: cppb $}
\svnid{$Id: CCT.tex 119 2013-10-02 15:24:08Z cppb $}

\makeatletter
\def\subsubsection{\@startsection{subsubsection}{3}{10pt}{-1.25ex plus -1ex minus -.1ex}{0ex plus 0ex}{\normalsize\bf}}
\def\paragraph{\@startsection{paragraph}{4}{10pt}{-1.25ex plus -1ex minus -.1ex}{0ex plus 0ex}{\normalsize\textit}}
\renewcommand\@biblabel[1]{#1}
\renewcommand\@makefntext[1]%
{\noindent\makebox[0pt][r]{\@thefnmark\,}#1}
\makeatother
\renewcommand{\figurename}{\small{Fig.}~}
\sectionfont{\large}
\subsectionfont{\normalsize}

\fancyfoot{}
%\fancyfoot[LO,RE]{\vspace{-7pt}\includegraphics[height=9pt]{headers/LF.pdf}}
%\fancyfoot[CO]{\vspace{-7.2pt}\hspace{12.2cm}\includegraphics{headers/RF.pdf}}
%\fancyfoot[CE]{\vspace{-7.5pt}\hspace{-13.5cm}\includegraphics{headers/RF.pdf}}
\fancyfoot[RO]{\footnotesize{\sffamily{1--\pageref{LastPage} ~\textbar  \hspace{2pt}\thepage}}}
\fancyfoot[LE]{\footnotesize{\sffamily{\thepage~\textbar\hspace{16.45cm} 1--\pageref{LastPage}}}}
\fancyhead{}
\renewcommand{\headrulewidth}{1pt}
\renewcommand{\footrulewidth}{1pt}
\setlength{\arrayrulewidth}{1pt}
\setlength{\columnsep}{6.5mm}
\setlength\bibsep{1pt}

\twocolumn[
  \begin{@twocolumnfalse}
\noindent\LARGE{\textbf{Counterion Condensation on Spheres in the Salt-free Limit$^\dag$}}
\vspace{0.6cm}

\noindent\large{\textbf{David A. J. Gillespie\textit{$^{a}$}, James E. Hallett\textit{$^{a,b}$}, Oluwapemi Elujoba\textit{$^{a}$}, Anis Fazila Che Hamzah\textit{$^{a}$}, 
Robert M. Richardson\textit{$^{b}$},  and Paul Bartlett$^{\ast}$\textit{$^{a}$}}}\vspace{0.5cm}
%Please note that \ast indicates the corresponding author(s) but no footnote text is required.

%\noindent\textit{\small{\textbf{Received Xth XXXXXXXXXX 20XX, Accepted Xth XXXXXXXXX 20XX\newline
%First published on the web Xth XXXXXXXXXX 20XX}}}
%
%\noindent \textbf{\small{DOI: 10.1039/b000000x}}
\vspace{0.6cm}
%Please do not change this text.

% % % % % % % % % % % % % % % % % % % % % % % % % % % % % % % % % % % % % % % % % % % % % % % % % % % % %
\noindent \normalsize{A highly-charged spherical colloid in a salt-free environment exerts such a powerful attraction on its counterions that a certain fraction condenses onto the surface of a particle. The degree of condensation depends on the curvature of the surface. So, for instance, condensation is triggered on a highly-charged sphere only if the radius exceeds a certain critical radius $\collrad^{*}$.  $\collrad^{*}$ is expected to be a simple function of the volume fraction of particles. To test these predictions, we prepare spherical particles which contain a covalently-bound ionic liquid, which is engineered to dissociate efficiently in a low-dielectric medium. By varying the proportion of ionic liquid to monomer we synthesise nonpolar dispersions of highly-charged spheres which contain essentially no free co-ions.  The only ions in the system are counterions generated by the dissociation of surface-bound groups. We study the electrophoretic mobility of this salt-free system as a function of the colloid volume fraction, the particle radius, and the bare charge density and find evidence for extensive counterion condensation. At low electric fields, we observe excellent agreement with Poisson-Boltzmann predictions for counterion condensation on spheres. At high electric fields however, where ion advection is dominant, the electrophoretic mobility is enhanced significantly which we attribute to hydrodynamic stripping of the condensed layer of counterions from the surface of the particle. }
% % % % % % % % % % % % % % % % % % % % % % % % % % % % % % % % % % % % % % % % % % % % % % % % % % % % %
\vspace{0.5cm}
%\\Generated: \currenttime, \today \\
%Revision: \svnrev \space  Saved: \svndate \space Author:  \svnFullAuthor{\svnauthor} \\
 \end{@twocolumnfalse}
  ]

\footnotetext{\textit{$^{a}$School of Chemistry, University of Bristol, Bristol BS8
1TS, UK.}}
\footnotetext{\textit{$^{b}$H.H. Wills Physics Laboratory, University of Bristol, BS8 1TL, UK.}}

\onecolumn % use for preprint
%\doublespacing % *************** use for review version
%\linenumbers

\twocolumn % use for archive version

% % % % % % % % % % % % % % % % % % % % % % % % % % % % % % % % % % % % % % % % % % % % % % % % % % % % %
\section{Introduction} \label{sec-intro}
% % % % % % % % % % % % % % % % % % % % % % % % % % % % % % % % % % % % % % % % % % % % % % % % % % % % %
%

The condensation of counterions plays a prominent role in a wide range of electrostatic soft matter, controlling not only the stability of colloids\cite{3206,4622} but also the compaction of genetic material\cite{Bloomfield1997}, %the folding of proteins, the adsorption of ions onto lipid membranes,
 and the self assembly of biomolecules such as actin and microtubules\cite{Wong2010}. The basic idea\cite{3206,4622} is that a highly charged object exerts such a long range attraction onto its counterions that a proportion condenses onto the surface effectively neutralising an equivalent amount of the structural charge $\Zstr$.  The charged substrate plus its captive counterions, may be considered as a single entity with an {\it effective} (or renormalized) charge $\Zeff$, which is significantly lower than the bare structural charge $\Zstr$. The difference $\Zstr-\Zeff$ is identified with the amount of counterions ``condensed'' onto the surface. Physically, ion condensation and the accompanying process of charge renormalization is driven by a competition between a favourable gain in electrostatic energy which occurs as ions collapse onto and neutralise the structural charge and an unfavourable loss of entropy when counterions bind to the surface. The process of counterion condensation occurs only in the limit of low salt concentration $\nsalt$ and hence long screening lengths, $\debye = (8 \pi \lB \nsalt)^{-1/2}$. Here $\lB = e^{2}/4 \pi \epsilon_{0} \epr \kBT$ is the Bjerrum length where $e$ is the electron charge, $\epsilon_{0}$ is the permittivity of free space,  $\epr$ the dielectric constant, and $\kBT$ the thermal energy.  While counterion condensation is a basic feature of highly charged matter the majority of experimental and theoretical attention to date has focussed on highly-charged rod-like polymers such as DNA where the phenomenon was first  analysed theoretically by Manning\cite{Manning1969a} and Oosawa\cite{Oosawa1971a}. Our understanding of charge condensation in non-cylindrical geometries, such as that onto a highly-charged sphere of radius $\collrad$, remains  quite rudimentary in comparison primarily because the long range character of Coulombic interactions in the low-salt regime where $\debye \gg \collrad$ ensures that the interactions are highly non-linear.

The importance of geometry on ion condensation see\-ms to have been explicitly discussed  first by Zimm and Le Bret\cite{Zimm1983} who analysed the ion distribution around cylinders, spheres and planes in the low-salt regime. Detailed numerical\cite{Deserno2000} and analytical studies\cite{Netz1998} of the mean-field Poisson-Boltzmann (PB) equation around a charged {\it cylinder} of radius $\collrad$, placed at the centre of a cylindrical Wigner-Seitz cell of radius $\RWS$, reveal that:
\begin{enumerate}[(i)]

\item  Counterion condensation is triggered only when the charge density exceeds a critical threshold value. For a cylinder with a vanishing radius, $\collrad \rightarrow 0$, condensation is observed when the Manning parameter $\Mann = \lB / \l$, a dimensionless measure of the linear charge density $e / \l$, exceeds unity.  

\item In the high charge limit ($\Mann > 1$), the number of counterions increases in proportion to the bare char\-ge density in such a way as to reduce the {\em effective} charge density to the critical value. Experimentally, this is reflected in an electrophoretic mobility $\mu$ which is independent of, or nearly so, of the linear charge density once the critical charge level $\Mann=1$ has been reached.

\item The proportion of counterions condensed onto the surface of the cylinder (equal to $1-1 / \Mann$) remains unchanged as the volume of the system is expanded indefinitely ($\RWS \rightarrow \infty$).

\end{enumerate}

The interaction of counterions with {\it spheres} differs from the 2D electrostatics of cylinders. Zimm and Le Bret\cite{Zimm1983} showed, for instance, that for a sphere in the salt-free limit  {\it all} of the bound counterions evaporate away on dilution, in contrast to (iii).  This difference reflects the dominance in the spherical geometry of the entropy of the counterions. The entropic contribution to the free energy varies only relatively slowly (logarithmically) in comparison to the electrostatic terms which depend on the inverse of the ion-sphere separation. Hence at large ion separations entropic  forces dominate. As a consequence there is no counterion condensation on an isolated sphere in the salt-free limit, where the particle is exposed only to solvent and an exact neutralizing number of counterions\cite{Zimm1983}. This result holds asymptotically for $\etac \rightarrow 0$, where $\etac = (\collrad / \WS)^{3} $ is the volume fraction of spheres.  

Ion condensation is still possible on spheres however at {\it finite} concentrations. In a remarkable series of papers\cite{Imai1952,Imai1953} begun in 1952 Imai and Oosawa demonstrated, from analytic properties of the spherical PB equation, the existence of a critical value of the particle charge $Z^{*}$  which separates two limiting cases. In the low charge limit ($Z \ll Z^{*}$) the diffuse cloud of counterions can be treated within linearized PB theory and there is no ion condensation. By contrast, in the high charge regime where $\Zstr > Z^{*}$,   counterion condensation occurs in the vicinity of the surface of the sphere. The critical particle charge $Z^{*}$ was found to be proportional to $\ln(1/\etac)$ so Zimm-Le Bret behaviour (no condensation) is recovered in the limit of vanishing concentration. Ohshima\cite{4516} later solved numerically the PB equation for a highly charged sphere in a salt-free environment and confirmed the earlier analytical predictions of Imai and Oosawa. Comparable conclusions are obtained from Oosawa's two-state model\cite{Oosawa1971a} in which the fraction of condensed counterions was optimized by a free energy minimization assuming coexistence between a condensed and a dilute gas-like phase of counterions\cite{Safran1990,Borukhov2004,Manning2007}. Finally, extensive numerical calculations\cite{4517,Ohshima2003,4616,Chiang2006a}  demonstrate that, in behaviour analogous to (ii), the electrophoretic mobility $\mu$ of a supercritical charged sphere ($\Zstr > Z^{*}$) is a constant, independent of the structural charge $\Zstr$. 

Given this large body of theoretical work it is perhaps surprising that from an experimental point of view, the behaviour of highly-charged spheres in the counterion-only limit remains largely unexplored (see however refs. \-\cite{4557,4623,4620,4424,Lobaskin2007a}). Most colloidal studies have focussed on the case of high salt concentrations and hence short screening len\-gths, $\kappa \collrad \gg 1$. Here the thickness of the condensed layer of counterions is small compared to the particle radius and the electrostatics are similar to those of a plane. Only when the sphere radius becomes comparable to the thickness of the condensate does this similarity break down and the effect of curvature is expected to become significant.  Consequently, to date many of the theoretical predictions on the influence of curvature on the extent of counterion condensation have not been extensively challenged by experiment.

In a recent paper\cite{Hussain2013}, we described the synthesis of a new class of highly-charged polymer particles which spontaneously charge in low-polarity solvents. The surface of a monodisperse particle is coated with ionic liquid groups which are engineered to efficiently dissociate in a nonpolar environment and to generate a high structural charge. In a nonpolar solvent such as dodecane, there is no equivalent to the auto-protolysis reaction of water to generate free charges, so the only ions in solution originate from the dissociation of surface groups. Consequently these particles provide an excellent experimental approximation to an idealized counterion only system. Low-polarity solvents offer further advantages for the study of counterion condensation. In a salt-free system the thickness of the condensed layer of ions is of order the Gouy-Chapman length\cite{Naji2005a} $\b = 1/(2 \pi \lB \sigma)$, where the surface charge density is $e \sigma$. The Bjerrum length $\lB$  is physically just the distance between two elementary charges which generates an electrostatic interaction of $\kBT$ so the maximum thermal charge density $\sigma$ is of order $\sim 1/\lB^{2}$. Consequently the Gouy-Chapman length $\b$ is expected to scale with the Bjerrum length $\lB$.   As a result the influence of curvature on ion condensation which occurs in aqueous systems on micellar length scales ($ \b \approx 1$ nm) \cite{4557,Manning2007,Carnal2012,Lamm2010}, is shifted in nonpolar solvents to the nanoparticle  size regime ($ \b \approx 50$ nm), where it may be more conveniently studied experimentally. Here, we study the electrophoretic mobility of charged nonpolar colloids in a salt-free suspension as a function of the colloid concentration, the radius of the particles, and their surface charge density. We interpret the measured mobility in terms of the effective particle charge and the degree of counterion condensation. Using numerical solutions of the non-linear PB equation, we compare our measurements against theoretical predictions and establish the general conditions defining the onset and extent of counterion condensation in dilute suspensions of highly charged spheres. The rest of this paper is organised as follows: in section~\ref{sec-exp} we describe our experimental system. The main features of theoretical models of counterion condensation are recalled in section~\ref{sec-theory}, while we report our experimental results in section~\ref{sec-results} before concluding in section~\ref{sec-conclude}.

% % % % % % % % % % % % % % % % % % % % % % % % % % % % % % % % % % % % % % % % % % % % % % % % % % % % %
\section{Materials and methods}  \label{sec-exp}
% % % % % % % % % % % % % % % % % % % % % % % % % % % % % % % % % % % % % % % % % % % % % % % % % % % % %

\subsection{Sample preparation}

\begin{figure*}[ht]
\centering
  \includegraphics[width=6cm]{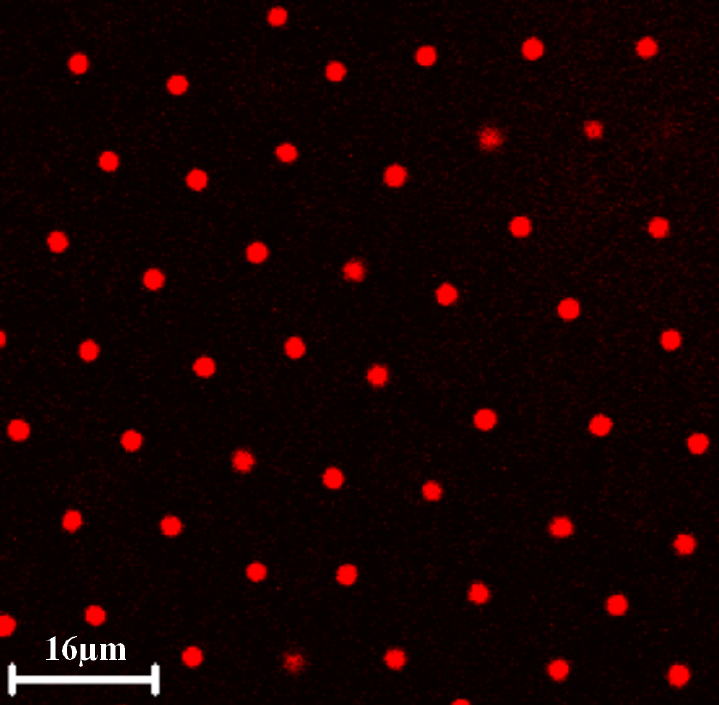}
  \caption{Charge generation in nonpolar suspensions. Confocal fluorescence microscopy image of the crystallization, in dodecane, of highly-charged poly(methyl methacrylate) particles (batch L12) as a consequence of long-range repulsive electrostatic interactions.}
  \label{fgr:confocal}
\end{figure*}

\begin{table}
	\centering
	\begin{threeparttable}[b]
	\caption{Charged nonpolar colloids}
	\label{tbl:samples}
	\begin{tabular}{lccccccccc}
	\toprule
	Batch & Ionic Monomer & $\W$ / Wt\% \tnote{a} & Monomer / Wt\% \tnote{b} & $\collrad$ / nm \tnote{c} & $\poly$ / \% \tnote{d}\\
	\midrule
	L1 & IM1 & 2.0 & 50 & 775 & 5 \\
	L2 & IM1 & 6.0 & 50 & 1265 & 9 \\
	L3 & IM2 &  2.0 & 50 & 440 & 8 \\
	L4 & IM4 & 1.4  & 45   &  490 & 6 \\
	L5 & IM2 & 4.0 & 24 & 38 & 7 \\
	L6 & IM2 & 8.0 & 25 & 46 & 7 \\
	L7 & IM2 & 11 & 23 & 56 & 6 \\
	L8 & IM3 & 2.0 & 50 & 300 & 4 \\
	L9 & IM3 & 3.9 & 50 & 434 & 7 \\
	L10 & IM3 & 5.9 & 50 & 442 & 8 \\
	L11 & IM2 & 4.0 & 45 & 145 & 11 \\
	L12 & IM2 & 6.0 & 50 & 950 & 4 \\
	\bottomrule
	\end{tabular}
	\begin{tablenotes}
	\footnotesize
	\item[a] Weight percentage of ionic monomer as a fraction of total monomer weight.
	\item[b] Weight percentage of MMA and MAA as a fraction of total preparation weight.
	\item[c] Average radius from dynamic light (DLS) and small-angle X-ray scattering (SAXS) measurements.
	\item[d] Coefficient of radius variation from electron microscopy and DLS.
	\end{tablenotes}
	\end{threeparttable}
\end{table}

The charged nonpolar system used consists of uniform spheres, comprising a copolymer of methyl methacrylate (MMA), methacrylic acid (MAA), and an ionic monomer (IM), dispersed in dried dodecane. The particle core is covered with a co\-valent\-ly-bound $\approx10$ nm thick outer polymeric shell of poly(12-hydroxy stearic acid-co-methyl methacrylate). The surface is functionalized with a number of highly hydrophobic ion-pairs formed from co\-valent\-ly-bound tetraalkly-ammonium cations and tetrakis[3,5-bis\-(tri\-fluoromethyl)phenyl]borate [TFPhB]$^{-}$  anions. The bulky size of the ions lowers the electrostatic cost of ionization so that a proportion of these ion-pairs dissociate in a low-polarity solvent\cite{Hussain2013}.  Accordingly the colloidal particles generate a spontaneous positive charge, which is counterbalanced by an equal number of free  [TFPhB]$^{-}$ counterions liberated into solution. Careful cleaning ensures that the counterion concentration exceeds the background ion concentration so that the colloids closely approximate a counterion-only system. 

	\begin{figure*}[h]
		\centering
			\includegraphics[width=10cm]{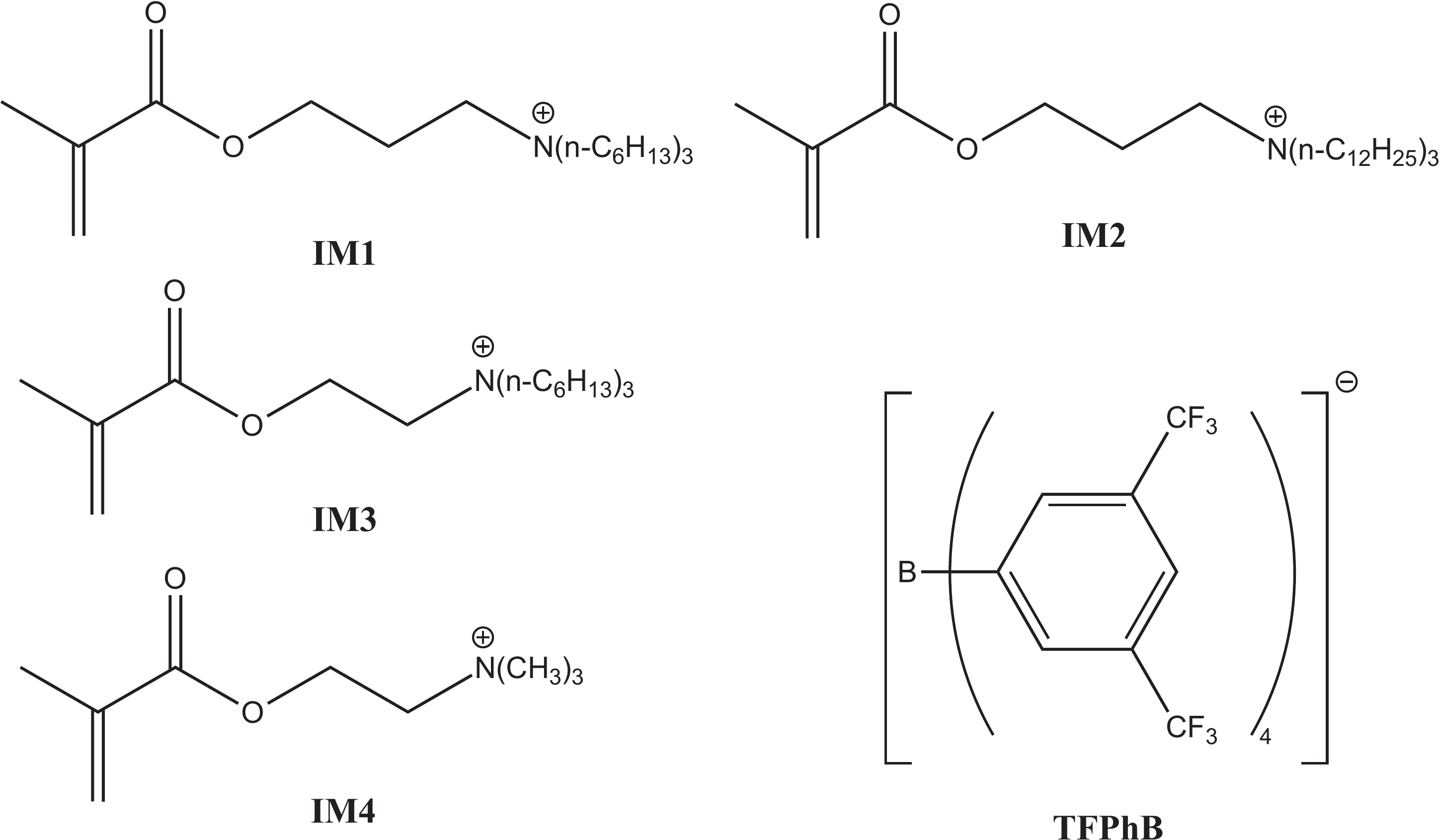}
			\caption{Ionic monomers consist of a tetrakis[3,5-bis(trifluoromethyl)phenyl]borate anion \textbf{TFPhB} and one of four quaternary ammonium cation: $n$-trihexyl-propyl-3-methacryloyloxy ammonium (\textbf{IM1}), $n$-tridodecyl-propyl-3-methacryloyloxy ammonium (\textbf{IM2}), $n$-trihexyl-ethyl-3-methacryloyloxy ammonium (\textbf{IM3}), or $n$-trimethyl-ethyl-3-methacryloyloxy ammonium  (\textbf{IM4}).}
			\label{fgr:PILs}
	\end{figure*}
	
At high concentrations the electrostatic repulsions between particles are large enough that the suspensions crystallize, as illustrated by the confocal image reproduced in Fig.~\ref{fgr:confocal}. In this work however we focus on the properties of fluid suspensions. Crystallization was avoided by working at low particle volume fractions in the range $2 \times 10^{-5} \leq \etac \leq 2 \times 10^{-3}$. The data presented here is obtained using twelve different batches of particles, labelled L1 -- L12 in Table~\ref{tbl:samples}, which were prepared by dispersion polymerization. The mean particle radius was varied from $\collrad = 38$ nm to  $\collrad = 1265$ nm by changing the concentration of MMA and MAA used in the synthesis. The particles obtained were spherical and had a high degree of size uniformity, as evidenced by a coefficient of radius variation, $C_{\mys{v}} = (\left< \collrad^{2}\right >- \left<\collrad \right>^{2})^{1/2} /  \left<\collrad \right>$, of between 4 \% and 11 \% where measured.  Four structurally different ionic monomers (IM1--IM4, Fig.~\ref{fgr:PILs}) were included at different concentrations in the particle synthesis to adjust the bare surface charge of the particles. All particles were studied in clean, dry dodecane  at room temperature. All colloids were purified by repeated cycles of centrifugation and redispersal in dry solvent to reduce the concentration of background ions. Mobility measurements indicate careful cleaning is the key to achieving high surface charge densities.

\subsection{Particle synthesis and purification} \label{sec:particle}

 Ionic liquid functionalized poly(methyl methacrylate) particles were prepared using the procedures described by Hussain et al.,\cite{Hussain2013}. Four different ionic monomers, prepared from the hydrophobic cations (IM1--IM4) and the bulky anion [TFPhB]$^{-}$, were synthesised. [Full details are contained within the Supplementary Information.] To remove unreacted ionic monomer and stray electrolyte from the samples, the particle suspensions were cleaned for up to twenty two repeated cycles of centrifugation and dispersal in fresh dodecane before use.  Dodecane was dried with the aid of activated 4 \AA{} molecular sieves. The conductivity of dried dodecane was $< 1$ pS cm$^{-1}$. The removal of stray electrolyte, introduced during the particle synthesis,  was monitored by measurement of the conductivity of the supernatant (Model 627 conductivity meter, Scientifica, UK) following each cycle of centrifugation. Purification was continued until the conductivity of the supernatant had dropped to $1.5 \pm 0.5$ pS cm$^{-1}$.

\subsection{Characterization methods}

The average radius $\collrad$ of the particles was determined from dynamic light scattering (DLS) measurements at 25\degree C in filtered dodecane. Measurements were performed  at $90\degree$ using a Malvern 4800 Autosizer (Malvern Instruments, UK) equipped with a 532~nm laser and a Malvern Zetasizer Nano S90 operating at 633~nm. Suspensions were diluted until the count rate recorded was $\approx 100 \times 10^{3}$ counts per second. The size polydispersity $C_{\mys{v}}$ was determined from scanning electron microscopy (SEM) for large particles ($\collrad > 250$ nm) and  DLS for smaller particles. SEM samples were prepared by diluting a drop of each dispersion in dodecane and then depositing onto a glass coverslip attached to an aluminium stub before sputter coating with gold. Images of about 1000 particles were collected and measured using image analysis software.  Confocal laser scanning microscopy of fluorescent labelled particles was performed using a Zeiss Pascal microscope. The fluorescent samples were excited using a 1 mW HeNe laser operating at 543 nm. Capillary tubes (ID 2.00 mm x 0.1 mm) were filled before being glued to glass slides and the ends sealed with epoxy resin.

Electrophoretic mobilities $\mu$ of the particles were determined using phase analysis light scattering. The technique involves driving charged particles with a low frequency AC electric field of magnitude $E$ through interference fringes generated by two intersecting laser beams.  The particle velocity $v$ is recorded by measuring the number of fringes a particle moves through. The  electrophoretic mobility was then calculated from the expression $\mu=v/E$. Experiments were performed at a scattering angle of $173\degree$ at 25\degree C  on a Malvern Zetasizer Nano ZS using a non-aqueous dip cell and a laser wavelength of 633~nm. Measurements were typically recorded at an applied field strength of $1.5 \times 10^{4}$ V m$^{-1}$. Systematic measurements of the variation of $\mu$ with $E$ (see Section~\ref{sec-field-effects}) showed that the mobility was constant for $E \leq 3 \times 10^{4}$ V m$^{-1}$ but increased at higher field strengths. All electrophoretic measurements, unless stated otherwise, were performed at a field of   $1.5 \times 10^{4}$ V m$^{-1}$ where the electrophoretic response was linear. Stock dispersions of known $\etac$ were prepared by dispersing an accurately weighed amount of PMMA particles (density $1.18$ g cm$^{-3}$) into dry dodecane (density 0.745 g cm$^{-3}$). Sequential dilutions were then performed using this stock to prepare a series of different $\etac$ samples.

% % % % % % % % % % % % % % % % % % % % % % % % % % % % % % % % % % % % % % % % % % % % % % % % % % % % %
\section{Counterion condensation} \label{sec-theory}
% % % % % % % % % % % % % % % % % % % % % % % % % % % % % % % % % % % % % % % % % % % % % % % % % % % % %

Counterion condensation on charged spheres under salt-free conditions has been studied theoretically by a number of authors\cite{Netz2003,Borukhov2004,Manning2007}. We briefly recall the main features of two popular models: the standard Poisson-Boltzmann (PB) model, which despite its drawbacks (such as the neglect of excluded-volume correlations betweens ions and colloid-colloid interactions), is a very good initial model in the weak coupling regime appropriate to our experiments; and second, the  Manning two-state model\cite{Borukhov2004,Manning2007} which has been very successful in predicting global properties, particularly for linear polyelectrolytes\cite{Manning1972}, and which has the advantage of numerical simplicity. We limit our discussion to the commonly-used cell model; in which the interactions within a dilute suspension of $N$ charged colloids in a volume $V$ is approximated by $N$ identical Wigner-Seitz cells each containing a single charged colloid placed at the centre of a spherical cell of radius $\WS$. Global charge neutrality is enforced by ensuring that the cell contains an appropriate number of counterions to exactly neutralise the charge on the central sphere. The volume of the cell is fixed by the available volume per colloid
\begin{equation}
\frac{4\pi}{3} \WS^{3} = \frac{V}{N}.
\end{equation}

\subsection{Electrostatic length scales}\label{sec:lengths}

\begin{table}
	\centering
	\begin{threeparttable}[b]
	\caption{Physical parameters for highly charged colloids in aqueous and nonpolar solvents}
	\label{tbl:length-scales}
	\begin{tabular}{l|ccccc|c}
	\toprule
	{\it Charged system} & Solvent   & $\collrad$ / [nm] & $|\sigma|$ / [$e/$nm$^{2}$]  & $\b$ / [nm] &  $\Xi$ & $\ms$ \\
	\midrule
	Poly(styrene) latex\cite{Jayasuriya1988a}          & H$_{2}$O    & 17  & 0.18    & 1.2 & 0.6 & 14 \\
	This work\tnote{a}           & C$_{12}$H$_{26}$ & 775 & 1.2 x 10$^{-4}$  & 46 &  0.6 & 17 \\
	\bottomrule
	\end{tabular}
	\begin{tablenotes}
	\footnotesize
	\item[] $\collrad$, and $\sigma$ are the radius and the surface charge density of the charged sphere. $\b$, $\Xi$, and  $\ms$ denote the Gouy-Chapman length, the Coulombic coupling constant, and the Manning radius, respectively. Monovalent ions are assumed. The Bjerrum length $\lB$ is taken as 0.71 nm for water (H$_{2}$O: $\epsilon = 78.4$), and 27.9 nm for dodecane (C$_{12}$H$_{26}$: $\epsilon = 2.01$) at 298 K.
	\item[a] Sample L1
	\end{tablenotes}
	\end{threeparttable}
\end{table}

A system of point-like counterions at a charged wall has two natural length scales\cite{Naji2005a}. Due to the dissociation of surface groups the surface acquires a homogeneous surface charge density of $e \sigma$. We assume for clarity that the  counterions are monovalent and negatively charged, so that $\sigma$ is positive by construction. Comparing the strength of the Coulombic interaction between counterions with the thermal energy $\kBT$ defines the Bjerrum length $\lB$, which is just the distance between two elementary charges which generates an electrostatic repulsion of $\kBT$. A second length characterizes the energy scale of the counterion-wall interaction. The Gouy-Chapman length $\b = 1/(2 \pi \lB \sigma)$ is the distance of a elementary charge from a uniformly-charged planar wall at which the wall-charge electrostatic interaction equals the thermal energy $\kBT$. Physically the Gouy-Chapman length is simply a measure of the thickness of the counterion layer formed at a planar wall. Indeed in PB theory, $\b$ equals the width of the layer  which contains exactly half of the total number of counterions\cite{Naji2005a}. The relative strengths of ion-ion and ion-surface correlations is embodied in  the {\it Coulombic coupling parameter} $\Xi$,
\begin{equation}
\Xi = \frac{\lB}{\b} = 2 \pi \lB^{2} \sigma.
\end{equation}
For large $\Xi$, ion-ion distances are large compared to the distance from the surface and the counterions form an essentially flat quasi-two dimensional layer on the charged wall. For small $\Xi$, on the other hand,  the ion-ion correlations are weak and fluid-like and the counterion distribution may be closely approximated by a mean-field description. Simulations indicate that the PB approximation is accurate provided that $\Xi < 1$, while the strong-coupling regime, where the PB analysis fails, occurs for $\Xi > 10$\cite{Moreira2001}. 

In a salt-free system of spheres, there exists a third independent length scale, the radius $\collrad$ of the charged particle. The free energy is a function of the electrostatic lengths, $\lB$, $\b$, and $\collrad$,  or since only the ratio between these quantities can be physically relevant, two dimensionless parameters. Rescaling all lengths in units of the Gouy-Chapman length defines the two characteristic ratios, $\Xi$ and the {\it Manning radius} $\ms$,
\begin{equation}
\ms = \frac{\collrad}{\b} = \frac{\Zstr \lB}{2 \collrad},
\label{eqn:manparam}
\end{equation}
where the last identity follows from the expression for $\sigma$ expressed in terms of the structural charge $\Zstr$ on a sphere, 
 \begin{equation}
\sigma = \frac{\Zstr}{4 \pi \collrad^{2}}.
 \end{equation}
The Manning radius $\ms$ is a dimensionless measure of the surface charge density of a  sphere and is the spherical analogue\cite{Netz2003} of the cylindrical charge parameter $\Mann$. 

In Table~\ref{tbl:length-scales} we compare the electrostatic length scales for a typical highly charged aqueous colloid (a 17 nm ion exchanged polystyrene latex\cite{Jayasuriya1988a}) and for the nonpolar sample L1 ($\collrad = 775$ nm) used in this work. As can be seen, the two systems have almost identical values of  $\Xi$ and $\ms$. However the Gouy-Chapman length is approximately forty times larger in the nonpolar system than the equivalent aqueous system. This means that it is possible to study the high curvature limit where $\collrad \simeq b$ using nonpolar colloids. In water this limit corresponds to radii of less than a few nm which is extremely difficult to probe with aqueous colloids.

\begin{figure*}[ht]
\centering
  \includegraphics[width=18cm]{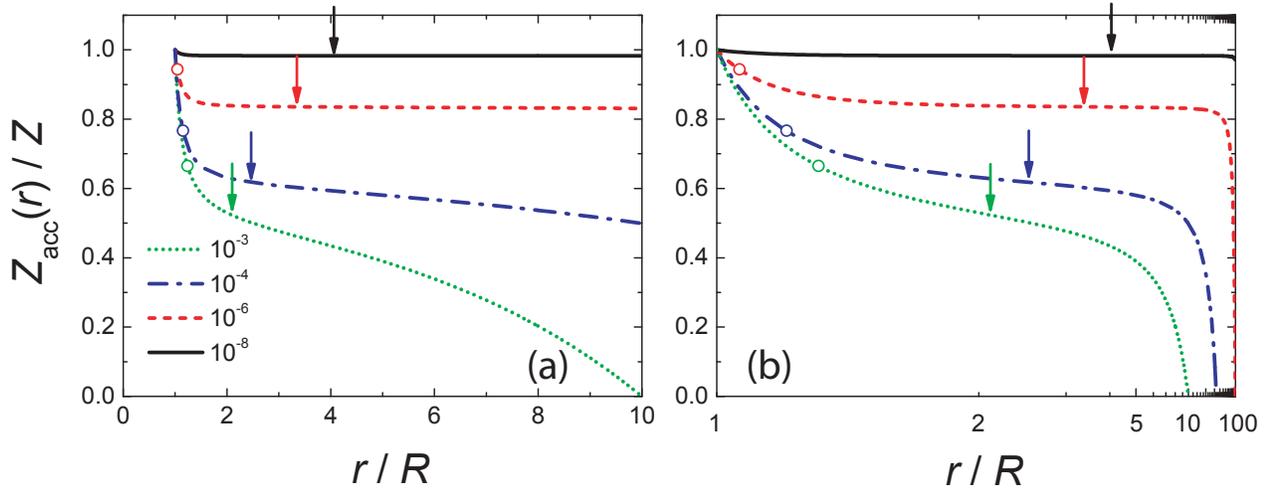}
  \caption{PB results for counterion condensation at particle volume fractions from bottom to top of $\etac \in \{10^{-3}, 10^{-4}, 10^{-6}, 10^{-8} \}$. The fraction $\Zacc(r)/\Zstr$ of total charge contained within a virtual sphere of radius $r$  around a charged particle, of radius $\ms = 8$, is plotted as a function of a linearly scaled $r$-axis in (a), and with a $1/r$ scaling in (b). The arrows mark the inflection points in the accumulated charge which define the location of the condensed layer of ions according to Belloni (Eq.~\ref{eqn-belloni}) while the circles indicate the predictions for the condensed layer proposed by Imai and Oosawa  (Eq.~\ref{eqn-imai-condition}).} 
  \label{fgr:PB-calcs}
\end{figure*}

\subsection{Poisson-Boltzmann model}

Within PB theory, the individual counterions are replaced by a spherically-symmetric density profile  $\rho(r) = \rho_{0} \mathrm{e}^{y(r)}$, which is solely a function of the radial coordinate $r$. In the absence of salt, the reduced electrostatic potential $y(r) = e \psi(r) /  \kBT $
satisfies the non-linear equation:
\begin{equation}\label{eqn-PB}
\nabla^{2} y = \frac{d^{2}y}{dr^{2}} + \frac{2}{r}  \frac{dy}{dr} = 4 \pi \lB \rho_{0} e^{y}
\end{equation}
where $\rho_{0}$ is an (arbitrary) reference concentration that fixes the zero of potential. The electrostatic boundary condition at the colloid surface arises from applying Gauss's law at $r=\collrad$,
\begin{equation}\label{eqn-BC1}
\left. \frac{\mathrm{d}y}{\mathrm{d}r} \right |_{r=\collrad} = - \frac{\sigma}{\epsilon_{0} \epr} = - \frac{2}{\b}
\end{equation}
where $\b$ is the Gouy-Chapman length. In the cell model, the electric field at the outer boundary of the cell vanishes as a result of charge neutrality so the second boundary condition is
\begin{equation} \label{eqn-BC2}
\left. \frac{\mathrm{d}y}{\mathrm{d}r} \right |_{r=\WS} = 0.
\end{equation}
Although the spherical PB equation (\ref{eqn-PB}--\ref{eqn-BC2}) can not be solved analytically it is easily solved numerically for the reduced potential $y(r)$, from which the counterion density $\rho(r)$ around an isolated colloid may be calculated. 

To  explore the ion distribution close to the colloidal surface, it is instructive to focus on the {\it total accumulated charge} $\Zacc(r)$, which is the charge (in units of $e$) found within a virtual sphere of radius $r \in [\collrad;\WS]$, and which from Gauss's law is related to the local electric field at $r$
\begin{eqnarray}
\frac{\Zacc(r)}{\Zstr} & = & 1 - 4 \pi \rho_{0} \int_{\collrad}^{r} \mathrm{e}^{y(r_{1})} r_{1}^{2} \mathrm{d}r_{1} \nonumber \\
 & = & \frac{\b}{2} \left (\frac{r}{\collrad} \right)^{2} \left | \frac{\mathrm{d}y}{\mathrm{d}r} \right |.
\end{eqnarray}
Since $y(r)<0$, $\Zacc(r)$ decreases  monotonically from $\Zacc(\collrad)=\Zstr$ at the surface of the particle to $\Zacc(\WS)=0$ at the edge of the cell. The latter result follows from charge neutrality (Eq.~\ref{eqn-BC2}). 
%
%In addition, we will also find it useful to use the related quantity
%\begin{equation}
%P(r) = 1- \frac{\Zacc(r)}{\Zstr}
%\end{equation}
%which is the fraction of ions contained within a sphere of radius $r$.

Figure~\ref{fgr:PB-calcs} shows the  accumulated charge around an isolated colloid with a high charge density corresponding to a Manning radius of $\ms = 8$, at a number of different volume fractions. The accumulated charge $\Zacc(r)$ displays a well-defined plateau which extends over a wide range of distances, when plotted as a function of  $r$,  as a consequence of the strong accumulation of counterions in the vicinity of the particle. This behaviour is particularly pronounced at low volume fractions where it validates the picture of a  particle surrounded by a thin layer of electrostatically-bound or {\it condensed} ions, with the remainder of the ions being located in an outer {\it diffuse} layer. Extensive theoretical work  has shown that,  at least for polyelectrolytes, the distinction between condensed and diffuse ions is in fact quite subtle\cite{Deserno2000} and a number of different criteria have been used to quantify the degree of counterion condensation. Probably the most reliable is a geometric criterion proposed by Belloni\cite{4622}, which for charged rods reproduces the condensed fraction of counterions predicted by Manning's and Oosawa's two-state model\cite{Deserno2000}. For a sphere,  Belloni identified condensed ions in terms of the inflection point in a plot of the accumulated charge $\Zacc$ as a function of the radial coordinate $1/r$. All ions within a distance  $\RI$ defined by the identity,
\begin{equation}\label{eqn-belloni}
 \left. \frac{\mathrm{d}^{2} \Zacc (r)}{\mathrm{d} (1/r)^{2}}  \right|_{r=\RI}  = 0
\end{equation}
are considered as condensed. Imai and Oosawa\cite{Imai1952,Imai1953} studied the analytical properties of the potential $y(r)$ in a salt-free suspension of charged spheres in the PB cell model. They showed that there is a qualitative change in the form of the potential at the point $r=\RF$ where the internal field  exceeds a critical threshold value
\begin{equation}\label{eqn-imai-condition}
% -\left. \frac{\mathrm{d}y}{\mathrm{d}r} \right|_{r=\RF} \geq \frac{3}{\collrad} \ln \left (\frac{\WS}{\collrad} \right)
 -\left. \frac{\mathrm{d}y}{\mathrm{d}r} \right|_{r=\RF} \geq \frac{1}{\collrad} \ln \left (\frac{1}{\etac} \right)
%  - \frac{\mathrm{d}y}{\mathrm{d}r}  >  \frac{1}{\collrad} \ln \left (\frac{1}{\etac} \right).
\end{equation}
so that all ions within a distance $\RF$ are regarded as condensed. As an illustration, Figure~\ref{fgr:PB-calcs} compares the inflection point rule  and the field based criterion for the identification of the condensed layer of ions. Note that the Imai-Oosawa criterion tends to underestimate the extent of condensation while the inflection point rule more accurately identifies the tightly bound layer of condensed ions found in the vicinity of the particle surface.

\subsection{Manning two-state model} \label{sec-manning}

The fraction of condensed ions can also be estimated analytically using a simple variational approximation. The counterions are split into two distinct sub-populations: condensed and free ions. The condensed fraction $\alpha$  is undetermined at the outset and is found by minimizing a free energy which encapsulates the competition between the favourable reduction in electrostatic energy and the unfavourable loss of entropy as ions bind to the particle. Assuming that a fraction $\alpha$ of the total number of counterions is condensed onto the surface of the sphere so that the effective particle charge is $\Zeff =(1-\alpha)\Zstr$ then the electrostatic charging energy $U_{\mys{el}}$  is
\begin{equation}
\frac{U_{\mys{el}}}{\kBT} = \frac{\lB (1-\alpha)^{2} \Zstr^{2}}{2\collrad}.
\end{equation}
The total free energy $F$  is a sum of $U_{\mys{el}}$ and a contribution from the translational entropies of the bound and free counterions.  The condensed ions are distributed within a thin shell around each sphere, whose thickness is comparable to the width $\b$ of the condensed layer on an infinite sheet. The volume available to the condensed ions is therefore of order $4 \pi \collrad^{2} \b$. The free ions occupy the remainder of the Wigner-Seitz cell, which in the limit $\etac \rightarrow 0$ is a region of finite volume $4 \pi \WS^{3} /3$. If the radius $a$ of the counterion is much smaller than the volume of the condensed region then the translational entropy of the ions is
\begin{equation}
-\frac{S_{\mys{ion}}}{\kB} = \alpha \Zstr \ln \left [ \frac{\alpha a^{3}}{3 \collrad^{2} \b}  \right ] + (1-\alpha)  \Zstr \ln \left [ \frac{(1-\alpha) a^{3}}{\WS^{3}}  \right ].
\end{equation}
Minimization of the free energy $F = U_{\mys{el}}-TS_{\mys{ion}}$ with respect to $\alpha$ yields the implicit equation for $\alpha$,
\begin{equation}\label{eqn:manning-fraction}
2 \ms (1-\alpha) = \ln \left ( \frac{\alpha}{1-\alpha}\right) + \ln \left ( \frac{\ms}{3 \etac} \right ),
\end{equation}
where $\ms = \collrad / \b$ is the Manning radius of the sphere. 

\subsection{Electrophoretic mobility}

Around a charged sphere  counterions accumulate in a thin shell. Seen from outside this shell, the effective charge on the sphere and its accompanying bound layer of counterions falls rapidly as the fraction of condensed ions rises. The effective charge may be determined experimentally from the electrophoretic mobility $\mu = v/E$, defined by the ratio of the particle drift velocity $v$ produced by an external electric field $E$. The Stokes mobility $\m0$ of a sphere of radius $\lB$ and charge $e$ provides a natural scale for the electrophoretic mobility in a solvent of viscosity $\eta$,
\begin{equation}
\m0 = \frac{e}{6 \pi \eta \lB}.
\end{equation}
An analytical expression for the electrophoretic mobility $\mu / \m0$ of a spherical particle in a salt-free suspension has been derived
by Ohshima\cite{Ohshima2003,4517} by approximating the exact electrokinetic equations of motion within a cell model. The mobility, at dilute particle concentrations ($\etac \ll 1$), is dependent on the product of the structural charge and the proportion of counterions located at the external cell boundary, namely
\begin{equation}\label{eqn-mu}
\frac{\mu}{\m0} = \left ( \frac{\Zstr \lB}{\collrad} \right ) \; \frac{\rho(\WS)}{\rho_{\mys{T}}} \; \Omega 
\end{equation}
where $\rho(\WS)$ is the concentration of counterions at the outer boundary of the cell $r=\WS$, $\rho_{\mys{T}}$  is the average counterion density in the cell $\rho_{\mys{T}} = 3\Zstr/4 \pi (\WS^{3}-\collrad^{3})$, and $\Omega$ is the volume-dependent term
\begin{equation}\label{eqn-omega}
  \Omega = 1 - \frac{9 \etac^{1/3}}{5} + \etac - \frac{\etac^{2}}{5}.
\end{equation}
In a cell model the osmotic pressure of the counterions is exactly given by $\kBT$ times the ion density at the outer cell boundary\cite{9473} so $\rho(\WS) / \rho_{\mys{T}}$ is simply the ratio of the osmotic pressure within the cell to the ideal osmotic pressure. At low charges, where the counterions are near-to-ideal then $\rho(\WS) / \rho_{\mys{T}} = 1$ and $\mu / \m0 \rightarrow \Zstr \lB / \collrad$ in the limit $\etac \rightarrow 0$. Hence the mobility ratio  $\mu / \m0$ at finite $\Zstr$ may be interpreted as a measure of the effective charge ratio $\Zeff \lB / \collrad$.

% % % % % % % % % % % % % % % % % % % % % % % % % % % % % % % % % % % % % % % % % % % % % % % % % % % % %
\section{Results and discussion} \label{sec-results}
% % % % % % % % % % % % % % % % % % % % % % % % % % % % % % % % % % % % % % % % % % % % % % % % % % % % %

\subsection{Existence of a  maximum mobility}

\begin{figure*}[ht]
\centering
  \includegraphics[width=14cm]{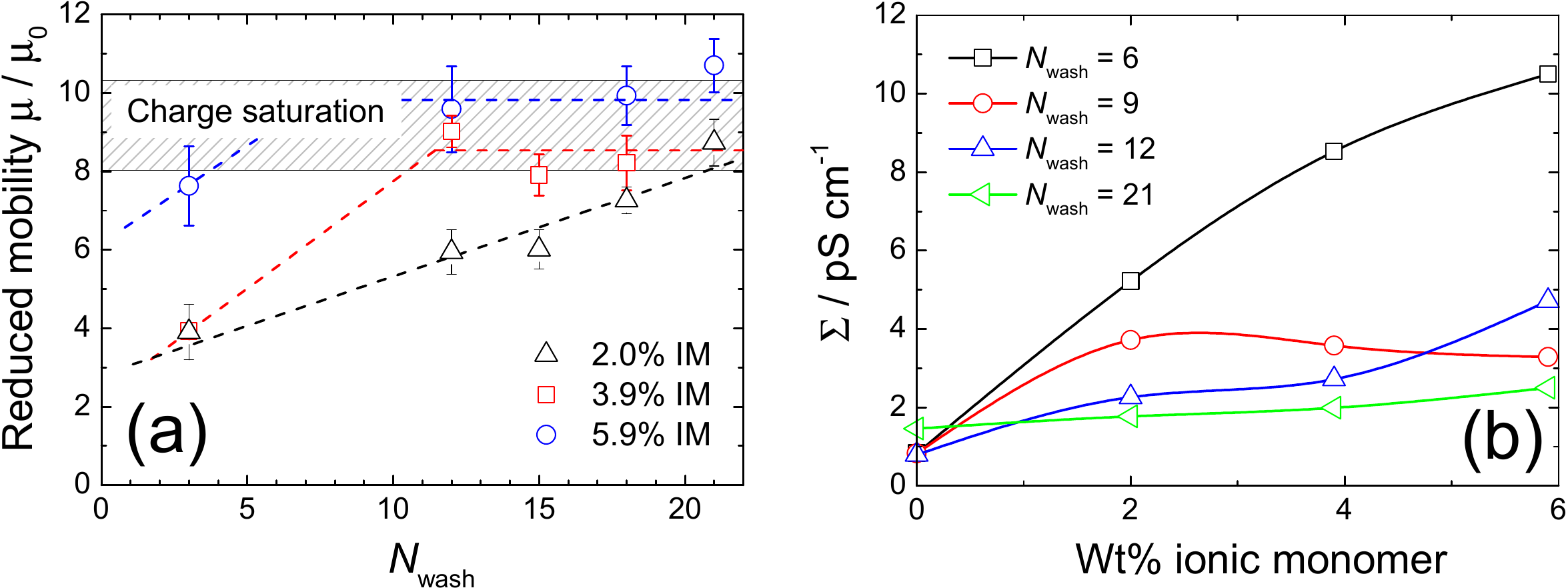}
  \caption{(a) Reduced electrophoretic mobility $\mu / \m0$ at $\etac = 6.3 \times 10^{-5}$ for batches L8--L10 in dodecane as a function of the number of centrifugation-redispersal cycles $N_{\mys{wash}}$. With cleaning, the mobility of particles initially increases before finally saturating at a plateau value (shown hashed in (a)). (b) Conductivity $\Sigma$ of supernatant after centrifugation of suspension as a function of the weight \% of ionic monomer incorporated. }
  \label{fgr:charge}
\end{figure*}

 A distinctive signature of counterion condensation is the existence of charge saturation, whereby as the bare charge $\Zstr$ grows the effective charge $\Zeff$ increases up to a constant plateau of $\Zeff = \Zeff^{*}$.  For an infinitely long charged cylinder the effective linear charge density $\Mann$, for instance, saturates at the critical value $\Mann^{*} = 1$, above which the effective charge density remains constant independent of the structural charge density\cite{Ramanathan1983}.  Similarly, in the case of an isolated charged sphere in the limit of small $\kappa \collrad$   Ramanathan\cite{Ramanathan1988} has shown that the effective charge of the particle reaches a maximum of
 \begin{equation}\label{eqn-Raman}
 \frac{\Zeff^{*} \lB }{\collrad} = -2 \ln \kappa \collrad -2 \ln (\ln \kappa \collrad) + 4 \ln 2 + O(1).
 \end{equation}
Charge saturation is revealed experimentally by the existence of a  maximum mobility, above which the mobility becomes independent of the bare  charge density.

To test for charge saturation in the salt-free limit  we synthesised a series of large nonpolar particles with different bare surface charge densities, batches L8--L10, containing 2.0, 3.9, and 5.9 \% by weight respectively  of the ionic monomer {\bf IM3}. Unreacted ionic impurities were removed by repeated cycles of centrifugation and redispersal in clean, dried dodecane. After each three wash cycles, the concentration of  particles was adjusted to a fixed volume fraction of $\etac = 6.3 \times 10^{-5}$ and the electrophoretic mobility $\mu$ and conductivity $\Sigma$ recorded. The resulting mobility data is shown in Fig.~\ref{fgr:charge}(a) as a function of the number of wash cycles $N_{\mathrm{wash}}$, for $N_{\mathrm{wash}} \leq 21$. Initially we find, as expected, that the mobility reflects the concentration of charged surface groups with the particles containing 5.9 \% of the ionic monomer displaying the largest mobility. However with further purification we identify two trends: first, the reduced mobilities  $\mu / \m0$ of the particles increases with $N_{\mathrm{wash}}$,  before finally  the electrophoretic mobilities of all particles saturate at a maximum $\mu / \m0 = 9.2 \pm 1.0$ (shown hashed in Fig.~\ref{fgr:charge}(a)). To rationalise these changes we note that Fig.~\ref{fgr:charge}(b) reveals that the increased mobility is accompanied by a significant drop in the conductivity and hence the concentration of free ions in suspension. Previous work\cite{Hussain2013} has indicated that the particle charge is controlled by the dissociation of ionic groups on the surface of the particle. The surface groups, which consist of bound tetraalkylammonium [TAA]$_{\mys{s}}$ cations and mobile [TFPhB] anions, are in dynamic equilibrium with any free ions in solution,
\begin{equation}
[\mathrm{TAA}][\mathrm{TFPhB}]_\mys{s} \rightleftharpoons [\mathrm{TAA}]_\mys{s}^{+} + [\mathrm{TFPhB}]^{-}
\label{eq:surface equilibrium}
\end{equation}
where the subscript s denotes a surface bound species. A high concentration of ionic impurities such as  $[\mathrm{TFPhB}]^{-}$ will act to shift the position of equilibrium back towards the left hand side and hence reduce the surface charge. Conversely, the reduction in the excess ion concentration which occurs after washing should lead to a progressive increase in the surface charge density of the particles. The data in Fig.~\ref{fgr:charge}(a) suggests that the mobility, which is proportional to the effective charge $\Zeff$, systematically increases with the charge density up to a limit of  $\mu / \m0 = 9.2 \pm 1.0$ (for this specific particle concentration), above which charge condensation occurs and the mobility becomes insensitive to the surface charge. 

\begin{figure*}[ht]
\centering
  \includegraphics[width=10cm]{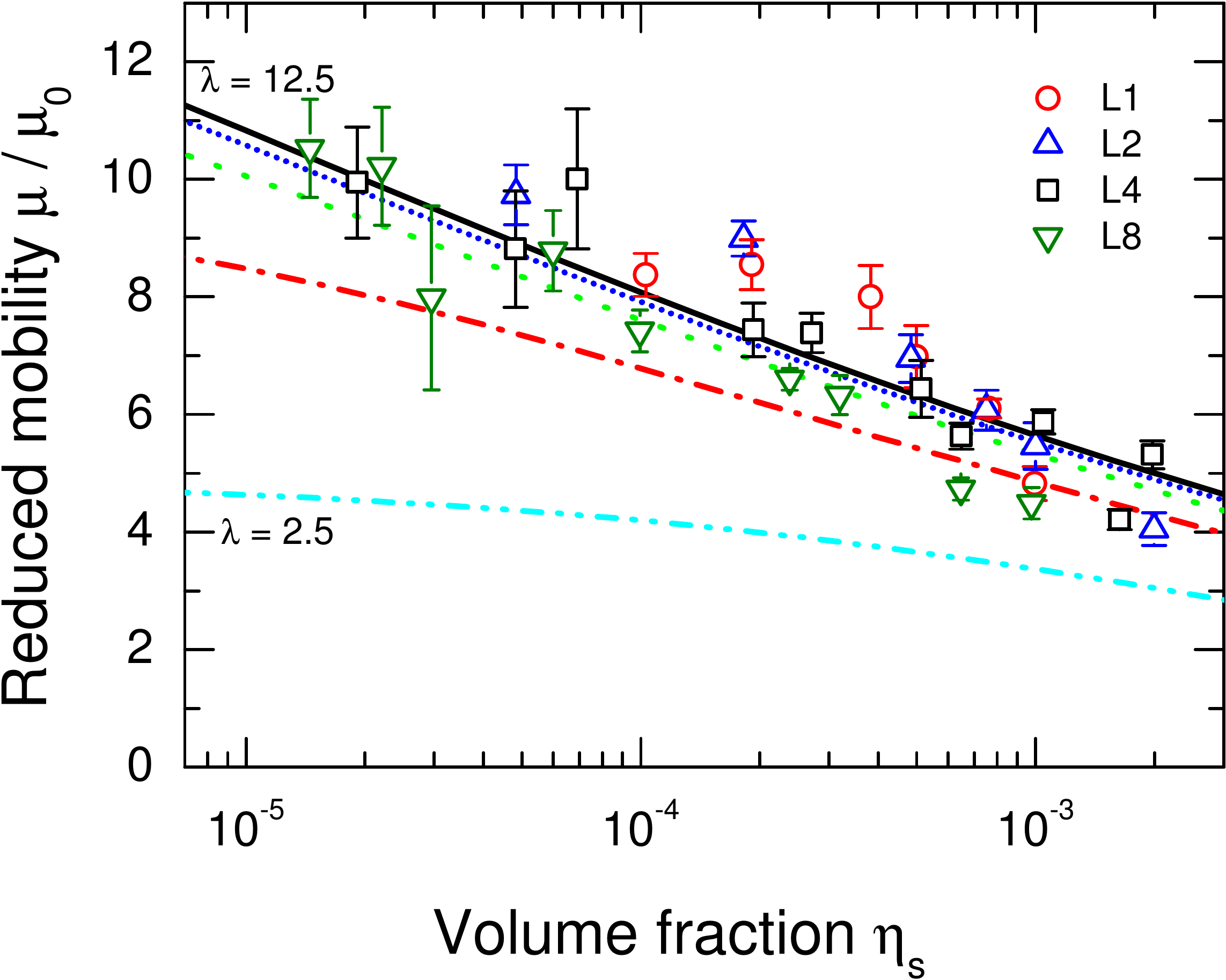}
  \caption{Reduced electrophoretic mobility $\mu / \m0$ as a function of colloid volume fraction $\etac$. The symbols denote different sample batches. Lines are predictions for scaled radii of $\ms = 2.5$ to $\ms = 12.5$, in steps of $\Delta \ms = 2.5$, from bottom to top. The predicted mobilities were obtained from numerical solutions of the full Poisson-Boltzmann equation and Eq.~\ref{eqn-mu}. Note the measured mobilities are independent of both the radius, ionic monomer type, and content and depend solely on the colloid volume fraction $\etac$.}
  \label{fgr:condensed}
\end{figure*}

\subsection{Counterion evaporation} \label{sec-condensation}

\begin{figure*}[ht]
\centering
\includegraphics[width=10cm]{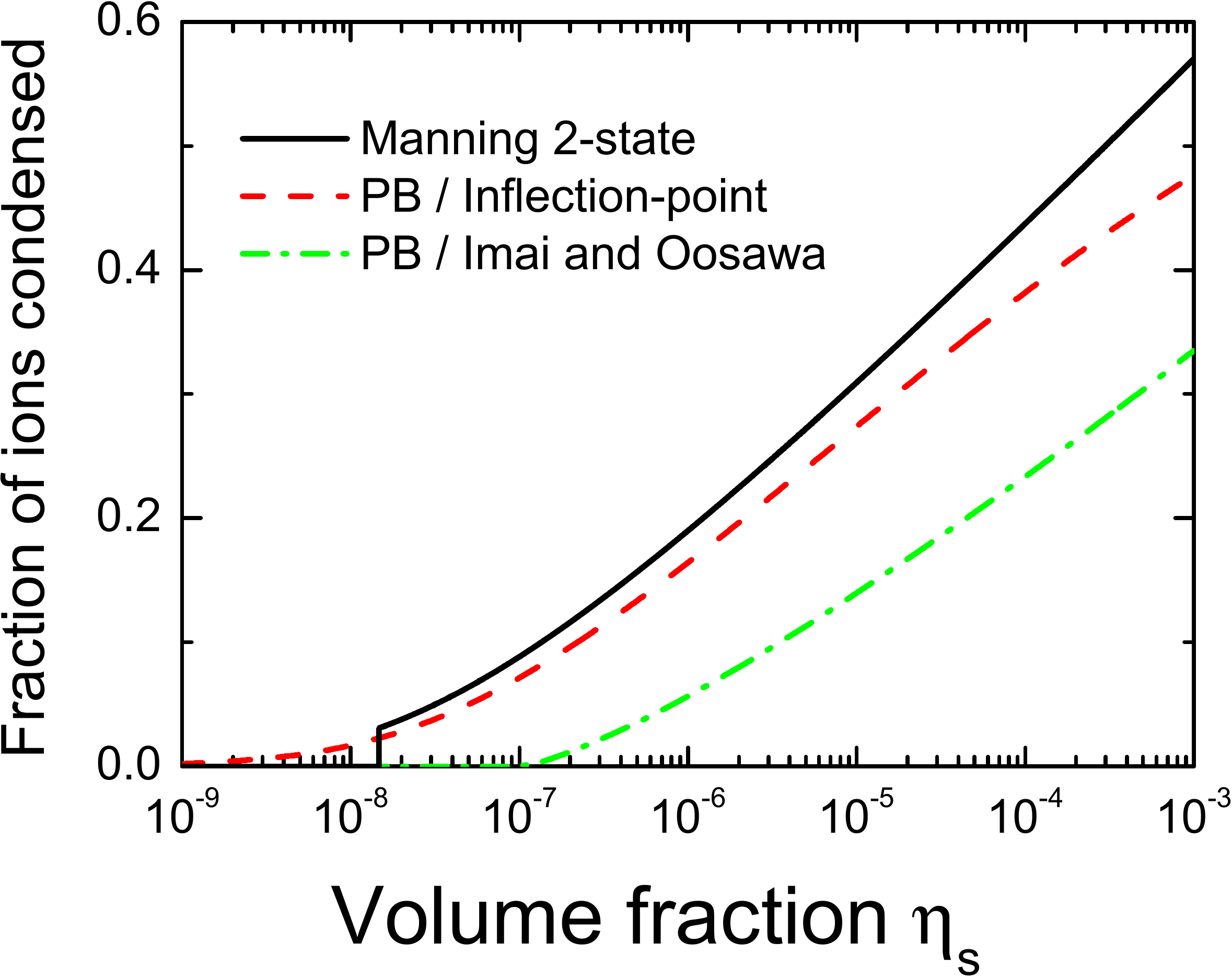}
  \caption{Counterion evaporation: comparison of predictions for the fraction of counterions condensed  onto a sphere with $\ms = 8$, as a function of the particle volume fraction $\etac$. The solid line is from the Manning two-state model (Eq.~\ref{eqn:manning-fraction}), while the dashed and dot-dashed lines are from Poisson-Boltzmann theory and the inflection-point criterion (Eq.~\ref{eqn-belloni}) or the Imai-Oosawa condition (Eq.~\ref{eqn-imai-condition}), respectively.}
  \label{fgr:condensed-fractions}
\end{figure*}

To confirm charge condensation we investigated a wide range of  cleaned particle suspensions of different radii and different ionic groups, at a variety of concentrations. The reduced mobilities  $\mu / \m0$ of particles synthesised using the monomers {\bf IM1}, {\bf IM3}, and {\bf IM4} are plotted in Figure~\ref{fgr:condensed} as a function of the particle concentration $\etac$. The data all overlay on a single master curve indicating that the particle radius, concentration of surface groups, and the type of these groups are unimportant in the condensed regime. Experimentally the mobility is purely a function of the particle concentration. Focussing, in particular, on the data for the two batches L1 (circles) and L2 (up triangles), which contain 2 \% and 6 \% respectively of the same ionic monomer, it is clear from Fig.~\ref{fgr:condensed} that the mobilities of the two samples are, to within experimental error, essentially indistinguishable over a wide range of volume fractions. Furthermore the lack, evident in Fig.~\ref{fgr:condensed}, of any correlation between the nature and concentration of surface groups and the measured particle mobilities highlights the insensitivity of the effective charge to the concentration or chemical structure of the ionic groups present at the surface.

On dilution, Figure~\ref{fgr:condensed} reveals that the maximum mobility $\mu$ of the counterion-only system increases sharply as $\etac \rightarrow  0$ with  $\mu$  growing approximately linearly with $\ln \etac$. This significant concentration dependence is very different from  that typically observed for suspensions in the salt-dominated regime ($\kappa \collrad \gg 1$) where the dependence of the electrophoretic mobility on the colloid volume fraction is weak. To understand the origin of this difference  we use the variational Manning model of a salt-free system (Sec.~\ref{sec-manning}).  The effective particle charge $\Zeff \lB/\collrad$ is $(1-\alpha) \Zstr \lB / \collrad$ which, in the limit of a high charge where $\lambda \gg 1$, may be estimated analytically from Eq.~\ref{eqn:manning-fraction} as,
\begin{eqnarray}
\Zeff \lB/\collrad & \approx &  \ln \left ( \frac{\ms}{3 \etac} \right ) \nonumber \\
					& \approx & \mathrm{constant} - \ln \etac. 
\end{eqnarray}
The increase in the effective charge $\Zeff \lB/\collrad$ with dilution stems from the dominance of the configurational entropy of the counterions over the electrostatic forces in dilute suspensions. As the average system volume per colloidal particle increases the layer of condensed ions  progressively evaporates  and the effective charge approaches the bare value, as predicted by Zimm and Le Bret\cite{Zimm1983}. 

Figure~\ref{fgr:condensed-fractions} shows predictions for the fraction of ions condensed for $\ms =8$ as a function of the particle concentration from the Manning model and the full PB theory, using both the inflection-point and the field-based criteria. While all three predictions are different in concept it is satisfying to see that the results are qualitatively very similar. Each of the three approaches reveal that the fraction of ions condensed on a sphere is density dependent, decreasing with a reduction in the particle volume fraction $\etac$. 
Quite remarkably, we find excellent agreement between experiment and a full numerical solution of the Poisson-Boltzmann equation using the cell model (Eq.~\ref{eqn-mu}). The predicted mobilities are shown by the lines in Figure~\ref{fgr:condensed}, for bare charge densities of $2.5 \leq \ms \leq 12.5$, in steps of $\Delta \ms = 2.5$. With an increase in the bare charge $\ms$ we see that the predicted mobilities converge rapidly to the charge saturated limit. The excellent agreement evident between experiment and theory is particularly encouraging because the comparison is totally free of any adjustable parameters and confirms the pivotal role of charge condensation in these nonpolar suspensions.

The monotonic decrease in the particle mobility seen with increasing $\etac$ is in sharp contrast with the dependence reported elsewhere for aqueous suspensions in the low-salt regime\cite{4623,4620,4424,Lobaskin2007a}, where regimes were observed in which the mobility ratio $\mu / \m0$ was either independent of $\etac$ or {\em increased} with increasing $\etac$. Although signal-to-noise considerations limit the lowest volume fraction which we can probe  to $\etac \gtrsim 10^{-5}$ we see no evidence for any behaviour other than an increase in mobility with decreasing $\etac$.  Consequently, it seems likely that the  behaviour reported is probably a consequence of the finite background ion concentration in aqueous systems, due to water dissociation.

\subsection{Counterion unbinding}

\begin{figure*}[ht]
\centering
  \includegraphics[width=10cm]{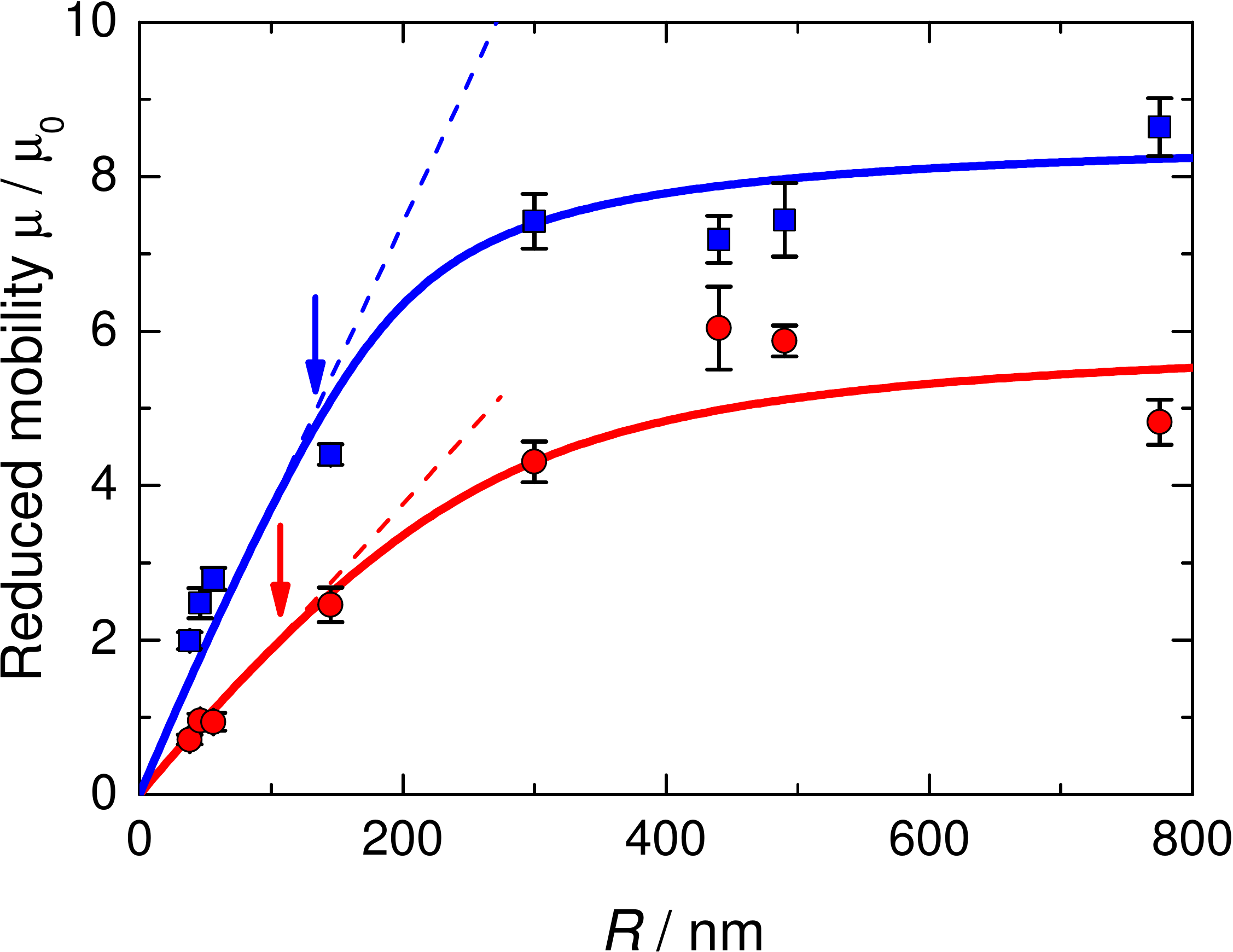}
  \caption{Counterion-unbinding transition. Reduced electrophoretic mobility $\mu / \m0$ as a function of particle radius $\collrad$ at fixed colloid packing fractions of $\etac = 10^{-4}$ (squares) and $\etac = 10^{-3}$ (circles). The solid curves represents a fit to the experimental data using a numerical solution of the PB equation and Eq.~\ref{eqn-mu} to calculate the particle mobility. The fitted values of the Gouy-Chapman length are $\b = 46.4 \pm 4.5$ nm at $\etac = 10^{-4}$ and $\b = 80.5 \pm 6.8$ nm at $\etac = 10^{-3}$.  The arrows mark the counterion condensation transition, from the Manning model and correspond well to the position at which the measured mobility deviates from the initial linear dependence on $\collrad$ (shown dashed).}
  \label{fgr:critical}
\end{figure*}

Ramanathan in a fascinating paper\cite{Ramanathan1988} written 25 years ago proved from the full PB equation that if the size of a highly charged sphere is reduced, keeping the surface charge density fixed, there is a critical size of the radius below which counterion condensation can not occur and the counterions must unbind from the surface. Ramanathan stated that the critical radius below which this unbinding transition occurred would be larger than the Bjerrum length but otherwise gave no explicit expression. While this striking prediction, which does not require any assumption beyond the validity of the Poisson Boltzmann model, has been checked using a field-theoretical formulation\cite{Netz2003} the result has not been confirmed by experiment. 

To establish if there is a critical size below which counterion condensation is absent the electrophoretic mobility was measured as a function of particle radius, at a fixed particle concentration. We vary the mean particle radius from $\collrad = 38$ nm  to 775  nm and monitored the mobility ratio $\mu / \m0 $ at fixed volume fractions $\etac = 10^{-4}$ and $10^{-3}$, as shown in Fig.~\ref{fgr:critical}. The largest particles display a substantial electrophoretic mobility which is independent of $\collrad$ and which agrees with the data plotted in Fig.~\ref{fgr:condensed} for systems with large degrees of ion condensation. In contrast Fig.~\ref{fgr:critical} reveals that the mobility of small particles, those with $\collrad \lesssim 300$ nm, is substantially lower than the largest particles with the mobility decreasing approximately linearly with decreasing $\collrad$. Equating the mobility ratio  $\mu / \m0$ with the effective charge $\Zeff \lB / \collrad$ this observation suggests that for small radii $\Zeff$ is quadratic in $\collrad$ and the charge density fixed, which implies the absence of counterion condensation.

\begin{figure*}[ht]
\centering
  \includegraphics[width=16cm]{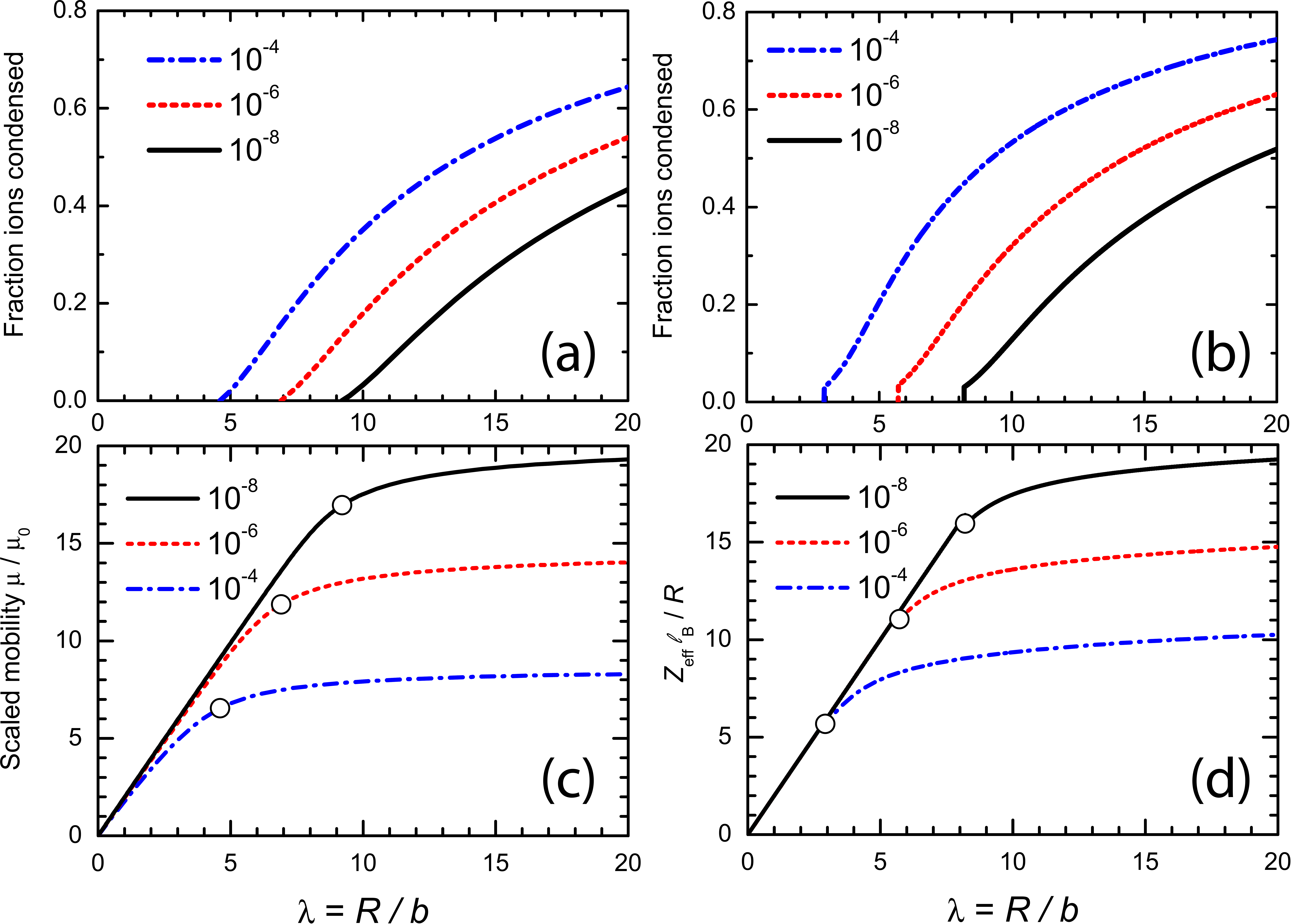}
  \caption{Typical numerical solutions of the PB equation (a) and the Manning two-state model (b) for the fractions of ions condensed as a function of the Manning radius $\ms$ for particle volume fractions of $\etac \in [10^{-4}, 10^{-6}, 10^{-8}]$ (from top to bottom). The PB calculations use the Imai-Oosawa  criterion (Eq.~\ref{eqn-imai-condition}) for condensation. The reduced electrophoretic mobility calculated from the PB solutions are plotted in (c), while the effective charge $\Zeff \lB / \collrad = (1-\alpha)\Zstr \lB / \collrad$ calculated from the Manning model is plotted in (d). The circles in (c) and (d) indicate the onset of counterion binding as the radius is increased.}
  \label{fgr:Manning-fractions}
\end{figure*}

To check this hypothesis we calculate the fraction of ions condensed and the corresponding electrophoretic mobilities from the full PB equation and the variational Manning two-state model. Figure~\ref{fgr:Manning-fractions} shows the predictions as a function of the scaled particle radius $\ms = \collrad / b$ in systems with volume fractions $\etac \in \{10^{-4},10^{-6},10^{-8}\}$.   While the two predictions for the size dependence of $\alpha$ differ quantitatively, with the Manning model consistently predicting a higher degree of ion condensation than the PB calculations, the shape of the two functions $\alpha(\ms)$ remain similar and show that the curvature of the particle plays a significant role in determining the degree of condensation. Below a critical radius $\ms < \ms^{*}$ no ion condensation occurs.  However as the sphere radius is increased above $\ms^{*}$ the degree of condensation becomes progressively stronger. The critical radius $\ms^{*}$ required for the onset of condensation increases as the volume fraction of particles is reduced, reflecting the dominance of the entropy of the counterions at low volume fractions. Physically the dependence on $\collrad$ is due to the increased distortion of the sheath of tightly-bound counterions when wrapped around a sphere compared to a flat plane. For a small radius a higher surface charge density is necessary to overcome the loss of electrostatic energy associated with this distortion. This curvature-driven unbinding transition is evident in the calculated mobilities shown in Fig.~\ref{fgr:Manning-fractions} (c) and (d). The variation of $\mu / \m0$ with $\ms$ may be divided conveniently into two regimes. At small $\ms$ the counterions are close to ideal and the osmotic pressure at the boundary of the Wigner-Seitz cell may be reasonably well approximated by the ideal gas expression, $ \rho_{\mys{T}} \kBT$. In this regime, from Eq.~\ref{eqn-mu}, the mobility increases linearly with $\ms$. However in highly charged systems where $\ms \gg 1$ the osmotic pressure at the cell boundary saturates as counterion condensation occurs near the surface of the particle and the mobility,  from Eq.~\ref{eqn-mu}, will approach a constant value, independent of the structural charge.  The circled points indicate the position of the critical radius $\ms^{*}$ where the PB and Manning models predicts the onset of condensation and the cross-over between the two radius-dependent regimes. 

To test these predictions, we compare the experimental values of the mobility ratio $\mu / \m0$ obtained at different $\collrad$ and $\etac$ with those calculated from the full PB theory using only the single fitting parameter $\b$. The agreement is excellent, as shown by the comparison between the data points and the lines in Fig.~\ref{fgr:critical} for suspensions with volume fractions $\etac = 10^{-4}$ and $10^{-3}$. The values of the fitted Gouy-Chapman length $\b$ are $46.4 \pm 4.5$ nm and $80.5 \pm 6.8$ nm respectively, which correspond to structural charge densities of $\sigma = 1.2 \pm 0.1 \times 10^{-4}$ nm$^{-2}$ and $0.071 \pm 0.006 \times 10^{-4}$ nm$^{-2}$. Interestingly, the value of the structural surface charge density is highest in the most dilute suspension which is consistent with an entropically-dominated charging mechanism\cite{3771}. Assuming the bulk entropy of the counterions exceeds the self-energy of a charged site on the particle surface, then we expect the surface charge density to be proportional to  $-\kBT \ln \etac$. The surface charge density at $\etac = 10^{-4}$ should then be approximately twice the value at $\etac = 10^{-3}$ which is in reasonable agreement with the charge density ratio determined experimentally of $1.73 \pm 0.22$.

\subsection{Counterion stripping} \label{sec-field-effects}

\begin{figure*}[ht]
\centering
 \includegraphics[width=10cm]{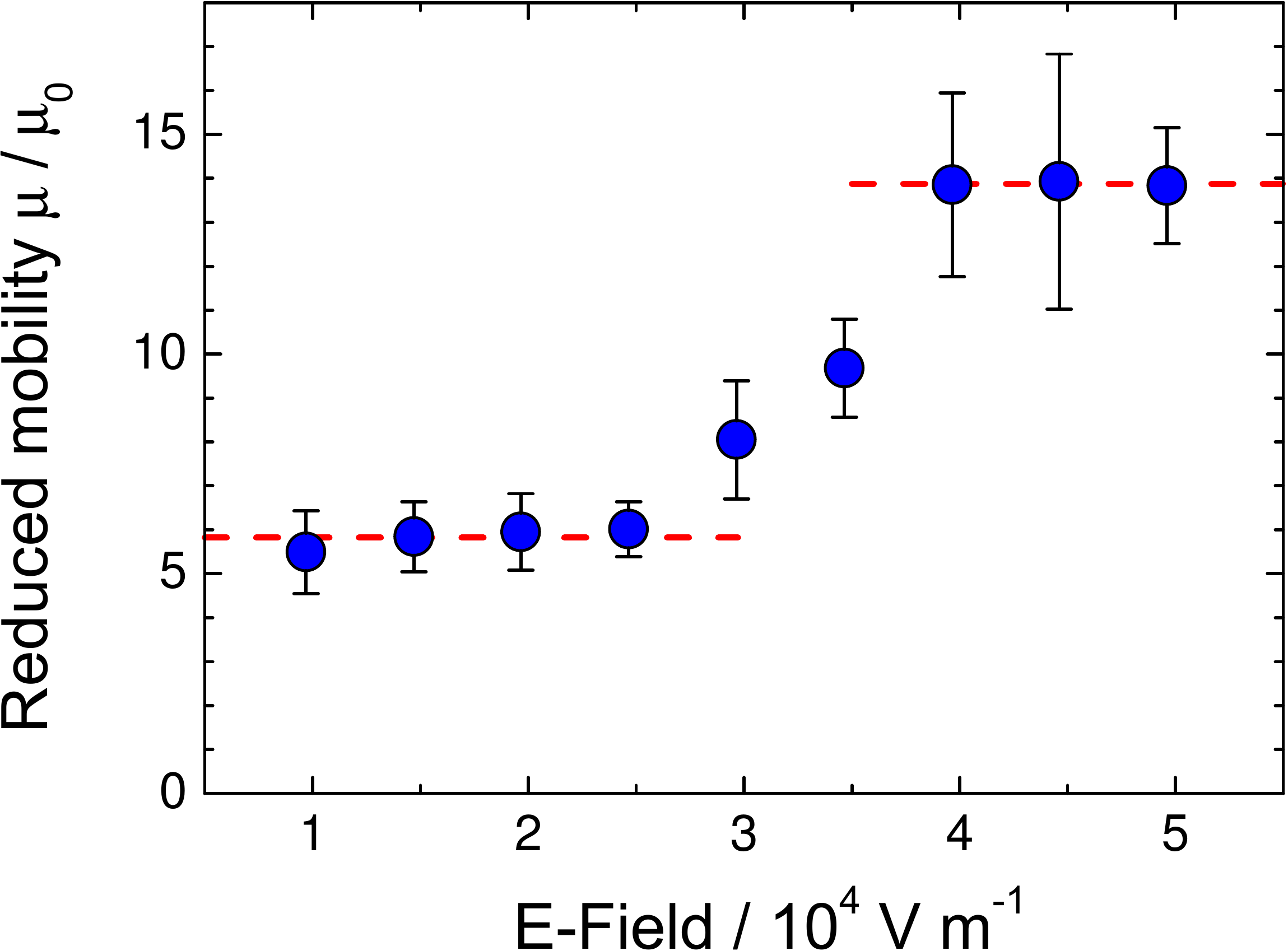}
  \caption{The field dependence of the reduced electrophoretic mobility measured on sample L1 ($\collrad = 775$ nm) at a fixed volume fraction of $\etac = 10^{-3}$.}
  \label{fgr:E-mobility}
\end{figure*}

The electrophoretic response of large charged particles ($\ms > \ms^{*}$) is suppressed as a result of strong counterion condensation. For instance, using the experimentally determined charge density at $\etac = 10^{-3}$, we estimate that the structural charge carried by a 775 nm radius sphere (batch L1) is of order $\Zstr \lB / \collrad = 19.3 \pm 2.5 $ while the relative mobility  is  a factor of more than three times smaller at $\mu / \m0 = 6$. This difference suggests that reducing the number of condensed ions should lead to a substantial enhancement in the response of a charged particle to an electric field. One way, potentially, to remove the tightly-bound layer of counterions is  by the application of a high electric field. In the resulting electrophoretic flow, the condensed layer will be distorted to a degree which is determined by a competition between the rate of ion diffusion and the rate at which ions are advected by the flow field. The resulting distortion is characterized by a dimensionless P\'{e}clet number, $\mathrm{Pe} = \collrad v / D_{\mys{i}}$, where $v$ is the electrophoretic flow velocity and $D_{\mys{i}}$ is the ion diffusion constant. The P\'{e}clet number quantifies the relative importance of flow advection and ion diffusion around a particle of radius $\collrad$. At small $\mathrm{Pe}$, ion diffusion is significantly faster than ion advection and there will be little distortion of the layer of condensed ions. At large $\mathrm{Pe}$ however, ion advection dominates and the counterions  are progressively stripped off the particle as it moves. In this nonlinear regime we expect a significant increase in the electrophoretic mobility as the effective charge rises towards the structural charge.  Recent lattice-Boltzmann simulations\cite{12476} of electrophoretic flows of highly charged colloids predict just such an enhancement of the mobility at P\'{e}clet numbers of order unity, the increase being particularly marked in conditions of low salt.  Nonpolar suspensions are ideal candidates to investigate experimentally the resulting distortion of the counterion layer because in low conductivity solvents strong electric fields can be applied without significant current flows  and the nanometer-sized ions in nonpolar solvents reduce the ion diffusion constant by some 1 or 2 orders of magnitude compared to aqueous systems so a wide range of $\mathrm{Pe}$ can be reached experimentally.

The marked nonlinear dependence of the electrophoretic mobility $\mu / \m0$ on the applied field $E$ is evident from the results shown in Fig.~\ref{fgr:E-mobility} for a suspension containing particles of batch L1 ($\collrad = 775$ nm) at a volume fraction of $\etac = 10^{-3}$. At low fields $E < 2.5 \times 10^{4}$ Vm$^{-1}$, which correspond to the small $\mathrm{Pe} < 0.16$ regime for nm-sized ions, the electrophoretic velocity $v$ is linear in the applied field and the mobility $\mu = v/E$ is  a constant, independent of $E$. All of the experimental data plotted in Figures~\ref{fgr:charge},~\ref{fgr:condensed}, and \ref{fgr:critical} are collected in this linear response regime, typically at $E = 1.5 \times 10^{4}$ Vm$^{-1}$ which corresponds to $\mathrm{Pe} = 0.09$. Figure~\ref{fgr:E-mobility} shows that increasing the field $E$  leads to a significant enhancement of $\mu$ with respect to these linear-response values. The mobility increases for $E > 3 \times 10^{4}$ Vm$^{-1}$ before appearing to saturate at the highest field, which correspond to $\mathrm{Pe} = 0.75$. The deviations from the linear response limit develops for $\mathrm{Pe} \simeq 0.25$, in reasonable agreement with simulation and theoretical predictions for the competing role of advection and diffusion\cite{12476}. The maximum mobility measured experimentally $\mu / \m0 \sim 14 \pm 2$ however seems to be slightly reduced in comparison to the structural charge of $\Zstr \lB / \collrad = 19.3 \pm 2.5 $ which may suggest that not all of the associated counterions have been stripped away from the moving particle. Understanding the physical mechanism in these counterion systems underlying the increase in the mobility $\mu$ with applied field  is an interesting open question which requires further investigation.

% % % % % % % % % % % % % % % % % % % % % % % % % % % % % % % % % % % % % % % % % % % % % % % % % % % % %
\section{Conclusions} \label{sec-conclude}
% % % % % % % % % % % % % % % % % % % % % % % % % % % % % % % % % % % % % % % % % % % % % % % % % % % % %

We have studied counterion condensation onto highly-charged spheres under conditions of no-added electrolyte, varying both the particle volume fraction, the radius, and the structural charge density. In contrast to much of the previous work in this area which has focussed on aqueous systems, we have used nonpolar colloids whose surfaces have been functionalized with ionic liquid groups to generate an appreciable charge density without any added electrolyte. Extensive purification ensures that the suspensions contains only the exact number of counterions needed to balance the charge on the particle. The electrostatic interactions between these particles and mobile counterions in low-dielectric solvents are surprisingly strong and long-ranged. 
While the functional dependence of the electrostatics is identical to that of charged particles in aqueous solvents the Gouy-Chapman length $\b$, which determines the thickness of the sheath of counterions at a flat charged surface, is approximately forty times larger than a comparable system in water.  By using electrophoresis we have been able to gain detailed information on the fraction of counterions tightly bound to the surface of a charged sphere. We have compared our experimental results where possible against numerical solutions of the non-linear Poisson-Boltzmann equation, and a simple variational model which yields analytic predictions. The form of the counterion density around a sphere is a function of the volume fraction $\etac$, the radius $\collrad$, and surface  charge density $\sigma$ of the particles. We identify two regimes which depend on the ratio of the radius of the sphere $\collrad$ to the Gouy-Chapman length $\b$. In the large radius (small curvature) limit we find that at high structural charge densities a layer of tightly-bound counterions is formed around a charged sphere. This layer reduces the effective charge $\Zeff$ down to a critical value at a fixed volume fraction, which is independent of the bare charge carried by the particle. The renormalized charge $\Zeff$ grows linearly with the radius $\collrad$ so in this regime the electrophoretic mobility $\mu$ is independent of $\collrad$.  Diluting the particle suspension causes the condensed layer to evaporate as the translational entropy of the counterions dominates the electrostatic interactions and the effective charge increases logarithmically with decreasing volume fraction $\etac$. In the small radius limit, where $\collrad \simeq \b$,  the curvature of the sphere is sufficiently high that the electrostatic attraction is no longer large enough to keep the counterions near the surface of the particle and there is an unbinding transition. At this point, the charge on the sphere is no longer renormalized and $\Zeff$ grows quadratically with the radius $\collrad$. In addition, we have shown that these systems display large non-linear electrophoretic responses. So, for instance, the electrophoretic mobility of large highly charged colloids is enhanced significantly in strong electric fields. 

Our results demonstrate that charged nonpolar suspensions are an excellent model system with which to study the equilibrium structure and non-equilibrium dynamics of counterion only systems, which have been studied extensively in the literature by both simulation and theory. Nonpolar colloids permit experiments to be performed and theory tested in regimes of electrostatics which are inaccessible to conventional aqueous systems.  Finally, our findings suggest new strategies for the optimization of the electrokinetic response of nonpolar suspensions which may find novel applications in electro\-phoretic displays and other nanodevices.

\section*{Acknowledgements}

Financial support from Merck Chemicals Ltd. UK, an affiliate of Merck 
KGaA, Darmstadt, Germany (DAJG)  is gratefully acknowledged. JEH is supported by EPSRC CDT grant EP/G036780/1. We thank the Diamond Light Source for X-ray beam time.


\balance

%\footnotesize{
%\bibliography{abbrev} %your .bib file
%\bibliographystyle{rsc} %the RSC's .bst file
%}

\providecommand*{\mcitethebibliography}{\thebibliography}
\csname @ifundefined\endcsname{endmcitethebibliography}
{\let\endmcitethebibliography\endthebibliography}{}

%\printnomenclature

\end{document}